\documentclass[aps,prd,12pt,nofootinbib,notitlepage]{revtex4-1}

\usepackage{amsfonts}
\usepackage{amsmath,epsfig,bm}
\usepackage{graphicx}
\usepackage{bm}
\usepackage{float}
\usepackage{color}

\usepackage{tabularx} 

\usepackage{times}

\newcommand{\be}{\begin{equation}}
\newcommand{\ee}{\end{equation}}
\newcommand{\ba}{\begin{eqnarray}}
\newcommand{\ea}{\end{eqnarray}}
\newcommand{\bd}{\begin{displaymath}}
\newcommand{\ed}{\end{displaymath}}

\def\thalf{{\textstyle{\frac{1}{2}}}}

\def\oneqt{{\textstyle{\frac{1}{4}}}}

\def\ones{{\textstyle{\frac{1}{6}}}}

\begin{document}

\title{{\bf Sphaleron Transition Rates and the Chiral Magnetic Effect}}
\author{Joseph I. Kapusta$^1$, Ermal Rrapaj$^{1,2}$, and Serge Rudaz$^1$}
\affiliation{$^1$School of Physics and Astronomy, University of Minnesota, Minneapolis, Minnesota 55455, USA \\
$^2$Department of Physics, University of California, Berkeley, California 94720, USA}

\vspace{.3cm}

\parindent=20pt

\begin{abstract}
The chiral magnetic effect is a novel quantum phenomenon proposed for
high-energy nuclear collisions but which has yet to be observed.  We
quantify the axial charge relaxation time, due to sphalerons, which enters in simulations
of this effect. An extrapolation of weak coupling calculations of the sphaleron rate yields
rather different relaxation times than strong coupling AdS/CFT calculations.
The AdS/CFT relaxation time is the larger one of the two by an order of magnitude, 
but the weak coupling relaxation time may not be reliable because it is only
marginally bigger than the microscopic thermalization time.  The role of quark masses
has yet to be accurately assessed.
\end{abstract}

\date{\today}

\maketitle


\section{Introduction}

The STAR Collaboration at the Relativistic Heavy Ion Collider (RHIC) measured the polarization of $\Lambda$ and $\bar{\Lambda}$ hyperons in Au+Au collisions over a wide range of beam energies \cite{FirstSTAR,Nature,SecondSTAR}.  These measurements are consistent with the suggestion that the large orbital angular momentum of the matter created in non-central heavy ion collisions could transfer to and polarize the quarks and subsequently the hadrons observed in the final state \cite{LiangWang1,Betz,Becattini1,Becattini2}.  The inferred vorticity $\omega \sim 10^{22}$ s$^{-1} \sim 6$ MeV is the highest ever observed for a fluid.  The observed hyperon polarization decreases with increasing beam energy, becoming nearly zero at the maximum RHIC energy of $\sqrt{s_{NN}} = 200$ GeV.   Measurements of the hyperon polarization in Pb-Pb collisions by the ALICE Collaboration at the much higher beam energies available at the Large Hadron Collider (LHC) are consistent with zero \cite{LambdaALICE}.  Vector mesons should also be polarized in non-central high energy heavy ion collisions \cite{LiangWang2,Becattini1,Becattini2,YangWang1,Tang}.  Somewhat surprisingly, spin alignment of the $K^{*0}(892)$ and $\phi (1020)$ vector mesons with the angular momentum was indeed observed by the ALICE Collaboration \cite{KphiALICE}.

The polarization measurements raise the important question of how long it takes for the strange quarks to reach and maintain equilibrium with the vorticity.  In Ref. \cite{KRR1} we studied several mechanisms, both of which lead to equilibration times three orders of magnitude too long to be relevant to heavy ion collisions.  In Ref. \cite{KRR2} we turned to a nonperturbative mechanism, namely the Nambu--Jona-Lasinio model for constituent quarks with the inclusion of the six-quark Kobayashi--Maskawa--'t Hooft interaction which breaks axial U(1)$_{A}$ symmetry.  This interaction can be viewed as arising from instantons whose presence rapidly decreases at temperatures $T$ above 150-200 MeV.  We found that constituent strange quarks do have time to equilibrate their spins with vorticity in that temperature range, just before hadronization occurs.  In Ref. \cite{KRR3} we showed that the thermal rate for helicity flip and for spin alignment with vorticity are equal, at least for $\omega \ll T$, which is the case in heavy ion collisions.  

Another interesting quantum phenomenon proposed for high energy heavy ion collisions is the chiral magnetic effect (CME) \cite{Fukushima1}.  This involves the appearance of an electric current and charge separation, in the direction of the large magnetic field generated by the colliding ions, due to a chiral imbalance.  This chiral imbalance originates in the axial anomaly.  It would be very exciting to observe the effects of nontrivial topological configurations of the gluon field in these collisions associated with the CME.  Readers are referred to review articles \cite{Kharzeev1,Kharzeev2} for more information.  However, to date there has been no convincing experimental evidence for the CME unlike the situation with spin polarization.  For updates see the proceedings of the Quark Matter Conferences \cite{QMseries}.

Instantons represent quantum mechanical tunneling through a barrier separating different topological sectors of QCD.  Sphalerons were first discovered in the electroweak interactions \cite{Klink} and baryon nonconservation in the early universe \cite{Kuz}.  They represent thermal fluctuations over the barrier and assume a high occupation of phase space by the bosonic fields so that classical equations of motion are an accurate representation of the physics.  That was followed by studies of strong interaction sphalerons relevant for high-energy nuclear collisions \cite{McLerran1,Mclerran2}.  They are responsible for relaxation to equilibrium when the axial charge is out of balance.  In this paper we focus our attention on the axial charge relaxation time.  In Sec. II we provide a brief review of how the axial charge equilibration time is obtained from the sphaleron solution at finite temperature.  In Sec. III we assemble previous results from the literature to calculate the axial charge equilibration time numerically as a function of temperature.  Questions addressed in this work  include  whether the strong coupling $\alpha_s$ is small enough for the sphaleron results to be valid, and  whether the kinetic thermalization time is small enough for the results to be trusted. We also discuss if the strange current quark mass $m_s$ is small enough that these results are applicable.  These are discussed in Secs. IV and V.  The summary and conclusions are presented in Sec. VI.

\section{Axial Charge and Sphalerons}

There is an explicit quantum breaking of the U(1)$_{A}$ symmetry in QCD resulting from the chiral anomaly 
\be
\partial_{\mu} j^{\mu}_{A} = \frac{\alpha_s N_f }{4 \pi} F_a^{\mu\nu} \tilde{F}^a_{\mu\nu} +2 i \sum_f m_f \bar{q}_f \gamma_5 q_f
\ee
where $j^{\mu}_{} = \sum_f \bar{q}_f \gamma_{\mu} \gamma_5 q_f$ involves a sum over quark flavors.  In addition chiral symmetry is explicitly broken by quark masses.  For now we assume that all $N_f$ quark flavors are massless.  Then chirality and helicity are the same and therefore so are the equilibration times.
There are several derivations of the axial charge rate equation which yield the equilibration time in the literature \cite{Khlebnikov,McLerran1,Moore1}.  Here we present a simple variation on those derivations.

From the divergence of the axial current the total charge in the system evolves with time as
\be
\frac{d Q_{A}(t)}{dt} = \frac{\alpha_s N_f }{4 \pi} \int d^3x F_a^{\mu\nu}(t,{\bf x}) \tilde{F}^a_{\mu\nu}(t,{\bf x})
\label{dQt}
\ee
assuming that the surface term is zero.  With no loss of generality we can choose an initial condition with zero net charge.  Then
\be
Q_{A}(t) = \frac{\alpha_s N_f }{4 \pi} \int_{-\infty}^t dt' \int d^3x' F_a^{\mu\nu}(t',{\bf x}') \tilde{F}^a_{\mu\nu}(t',{\bf x}')
\label{Qt}
\ee
We would like to do an averaging of Eq. (\ref{dQt}) in a thermal bath at temperature $T$ with the statistical density matrix
\be
\rho = \frac{1}{Z} \exp\left(-\beta H \right)
\ee
This means multiplying Eq. (\ref{dQt}) by $\rho$ and taking the trace.  

Now consider a small fluctuation in the axial chemical potential $\delta \mu_A(t)$.  This results in a change in energy $\delta H = Q_{A}(t) \delta \mu_A(t) $.  Expand the exponential in powers of $\beta Q_{A}(t) \delta \mu_A(t) $ with $Q_{A}(t)$ given by Eq. (\ref{Qt}).  The zero order term on the right side vanishes in the unperturbed ensemble because it has zero net axial charge.  The first order term gives
\be
\frac{d \langle Q_{A}(t) \rangle}{dt} = \left(\frac{\alpha_s N_f }{4 \pi}\right)^2 \int_{-\infty}^t dt' \int d^3x' \int d^3x 
\left\langle F_{a}^{\mu\nu}(t',{\bf x}') \tilde{F}^{a}_{\mu\nu}(t',{\bf x}')
F_b^{\kappa\sigma}(t,{\bf x}) \tilde{F}^b_{\kappa\sigma}(t,{\bf x}) \right\rangle \beta \delta \mu_A(t)
\label{avedQt}
\ee
where the angular brackets refer to averaging in the unperturbed ensemble.  The calculation is valid when the axial charge is small compared to the number of degrees of freedom, namely, $|Q_A| \ll s V$ where $s$ is the entropy density associated with the unperturbed density matrix and $V$ is the volume.  Because these are considered infinitesimal quantities, we can replace $\delta \mu_A(t)$ with $\langle Q_A(t) \rangle/V \chi_A$ where $\chi_A$ is the susceptibility.  Also, because the unperturbed ensemble is translationally invariant in both space and time 
\bd
\int_{-\infty}^t dt' \int d^3x' \int d^3x \left\langle f(t',{\bf x}') f(t,{\bf x})  \right\rangle
\ed
\bd
= V \int_{-\infty}^0 dt_{rel} \int d^3x_{rel} \left\langle f(t_{rel},{\bf x}_{rel}) f(0,{\bf 0})  \right\rangle
\ed
\be
= \thalf V \int_{-\infty}^{\infty} dt_{rel} \int d^3x_{rel} \left\langle f(t_{rel},{\bf x}_{rel}) f(0,{\bf 0})  \right\rangle
\ee
where in this case $f(t,{\bf x}) = F_{a}^{\mu\nu}(t,{\bf x}) \tilde{F}^{a}_{\mu\nu}(t,{\bf x})$.  

The sphaleron rate $\Gamma_{\rm sphal}$ is
\be
\Gamma_{\rm sphal} = \left(\frac{\alpha_s}{8 \pi}\right)^2
\int_{-\infty}^{\infty} dt_{rel} \int d^3x_{rel} \left\langle F_{a}^{\mu\nu}(t_{rel},{\bf x}_{rel}) \tilde{F}^{a}_{\mu\nu}(t_{rel},{\bf x}_{rel})
F_b^{\kappa\sigma}(0,{\bf 0}) \tilde{F}^b_{\kappa\sigma}(0,{\bf 0}) \right\rangle
\label{Srate}
\ee
This arises in the following way.  Conventionally it is defined with one flavor as
\be
\lim_{t \rightarrow \infty} \oneqt \left\langle \left( Q_{A}(t) - Q_{A}(0) \right)^2 \right\rangle = \Gamma_{\rm sphal} \, V t
\ee
which is just a random walk.  The factor of $(1/2)^2$ is due to the difference between the Chern-Simons number and the axial charge.  Inserting Eq. (\ref{Qt}) leads to  
\bd
\oneqt \left\langle \left( Q_{A}(t) - Q_{A}(0) \right)^2 \right\rangle = 
\ed
\bd
\left(\frac{\alpha_s}{8\pi}\right)^2
\int_0^t dt' \int d^3x' \int_0^t dt'' \int d^3x'' \left\langle F_{a}^{\mu\nu}(t',{\bf x}') \tilde{F}^{a}_{\mu\nu}(t',{\bf x}')
F_b^{\kappa\sigma}(t'',{\bf x}'') \tilde{F}^b_{\kappa\sigma}(t'',{\bf x}'') \right\rangle =
\ed
\be
V t \left(\frac{\alpha_s}{8\pi}\right)^2
\int_{-t/2}^{t/2} dt_{rel} \int d^3x_{rel} \left\langle F_{a}^{\mu\nu}(t_{rel},{\bf x}_{rel}) \tilde{F}^{a}_{\mu\nu}(t_{rel},{\bf x}_{rel})
F_b^{\kappa\sigma}(0,{\bf 0}) \tilde{F}^b_{\kappa\sigma}(0,{\bf 0}) \right\rangle
\ee
where space and time translation invariance in the thermal ensemble was used in the last line.

Using the free gas value for the susceptibility $\chi_A = N_f T^2$ gives the final differential equation
\be
\frac{d \langle Q_{A}(t) \rangle}{dt} = - 2 N_f \frac{\Gamma_{\rm sphal}}{T^3} \langle Q_{A}(t) \rangle
\ee
This means that any fluctuation in the axial charge will relax to its equilibrium value of zero with a time constant
\be
\frac{1}{\tau_A} = 2 N_f \frac{\Gamma_{\rm sphal}}{T^3}
\ee
This is basically an illustration of the fluctuation-dissipation theorem.

\section{Sphaleron Rate}

As pointed out in Ref. \cite{McLerran1} there is no finite energy classical solution in QCD corresponding to the sphaleron in the symmetry broken phase of electroweak theory.  Nevertheless the term sphaleron is used to describe topological transitions in QCD.  The most accurate calculation of the sphaleron transition rate for weak coupling we are aware of for the SU(3) gauge group was reported in Ref. \cite{Moore1}.  In the limit of weak coupling and high temperature \cite{Arnold1,Arnold2}
\be
\Gamma_{\rm sphal} = \frac{792}{6+N_f} \left[ \ln\left(\frac{m_D}{\gamma}\right) + 3.041 \right] \alpha_s^5 T^4
\label{weak}
\ee
with color Debye mass
\be
m_D^2 = \left( 1 + \ones N_f \right) 4\pi \alpha_s T^2
\ee
and rate of color randomization 
\be
\gamma = 3 \alpha_s T \left[ \ln\left(\frac{m_D}{\gamma}\right) + 3.041 \right]
\ee
determined self-consistently.  This semi-analytical rate was compared to numerical simulations of Hamiltonian dynamics in real time on a spatial lattice in Ref. \cite{Moore1}.  For $N_f = 3$ and $\alpha_s = 0.023$ the numerical result is consistent with the semi-analytical result within statistical uncertainties.  As $\alpha_s$ is increased the ratio of numerical to semi-analytical results becomes smaller, reaching $0.624 \pm 0.011$ at the largest value $\alpha_s = 0.093$ used in the calculations.  For larger values of $\alpha_s$ it became difficult to distinguish topological behavior on the lattice.  In their conclusions the authors point out that the Hamiltonian dynamics used is not a successful description of QCD even at temperatures as high as the electroweak scale.  

The AdS/CFT correspondence has been used to calculate the rate at strong 't Hooft coupling $g^2 N_c \gg 1$ in ${\cal N} = 4$ Super Yang-Mills theory.  It is \cite{SS}
\be
\Gamma_{\rm sphal} =  \frac{(\alpha_s N_c)^2}{16\pi} T^4
\label{AdSCFT}
\ee
This rate will be compared to the weak coupling and numerical results in the next section.

The sphaleron calculations are only valid when the microscopic equilibration time is small compared with $\tau_A$.  Parametrically, the former has the form $1/\alpha_s^2 T$ but the coefficient might matter numerically.  We take the numerical estimate from the calculations in Ref. \cite{Kurkela1}.  Using an effective kinetic theory they determine that the momentum thermalizes with a time constant 
\be
\frac{1}{\tau_{eq}} = \left[ 1 - 0.12 \ln(12 \pi \alpha_s) \right] 2 \pi^2 \alpha_s^2 T
\label{micro}
\ee
The logarithmic dependence comes from an infrared divergence in momentum diffusion caused by soft elastic collisions.  This function represents their numerical results very well in the range of couplings used, namely $5/600 \pi \le \alpha_s \le 5/6 \pi \approx 0.265$.

It should be emphasized that both equilibration times are computed with gauge field dynamics only.  Dynamical quarks are not included in either calculation.

To obtain numerical results we need to know $\alpha_s(T)$.  The highest order perturbative calculation of the pressure was compared to lattice QCD results in Ref. \cite{Albright1} using the $\beta$ function up to 3 loops.  See Eqs. (37) to (39) in Appendix B of that paper.  Those results should be good for $T > 180$ MeV.

\section{Numerical Results}

In this section we present numerical results for the strong coupling $\alpha_s$ as a function of temperature, the sphaleron rates, the ratio of kinetic to axial charge equilibration times, and for the absolute axial charge equilibration time.  We consider the temperature range between 200 and 1000 MeV, although the highest temperatures achieved in heavy ion collisions likely does not exceed 600 MeV.

Figure 1 shows $\alpha_s$ as a function of temperature for 3 flavors of quarks.  These are taken from fitting to lattice QCD with the physical quark masses in Ref. \cite{Albright1}.  The dependence is essentially logarithmic of course.

\begin{figure}[H]
\center{\includegraphics[width=290pt]{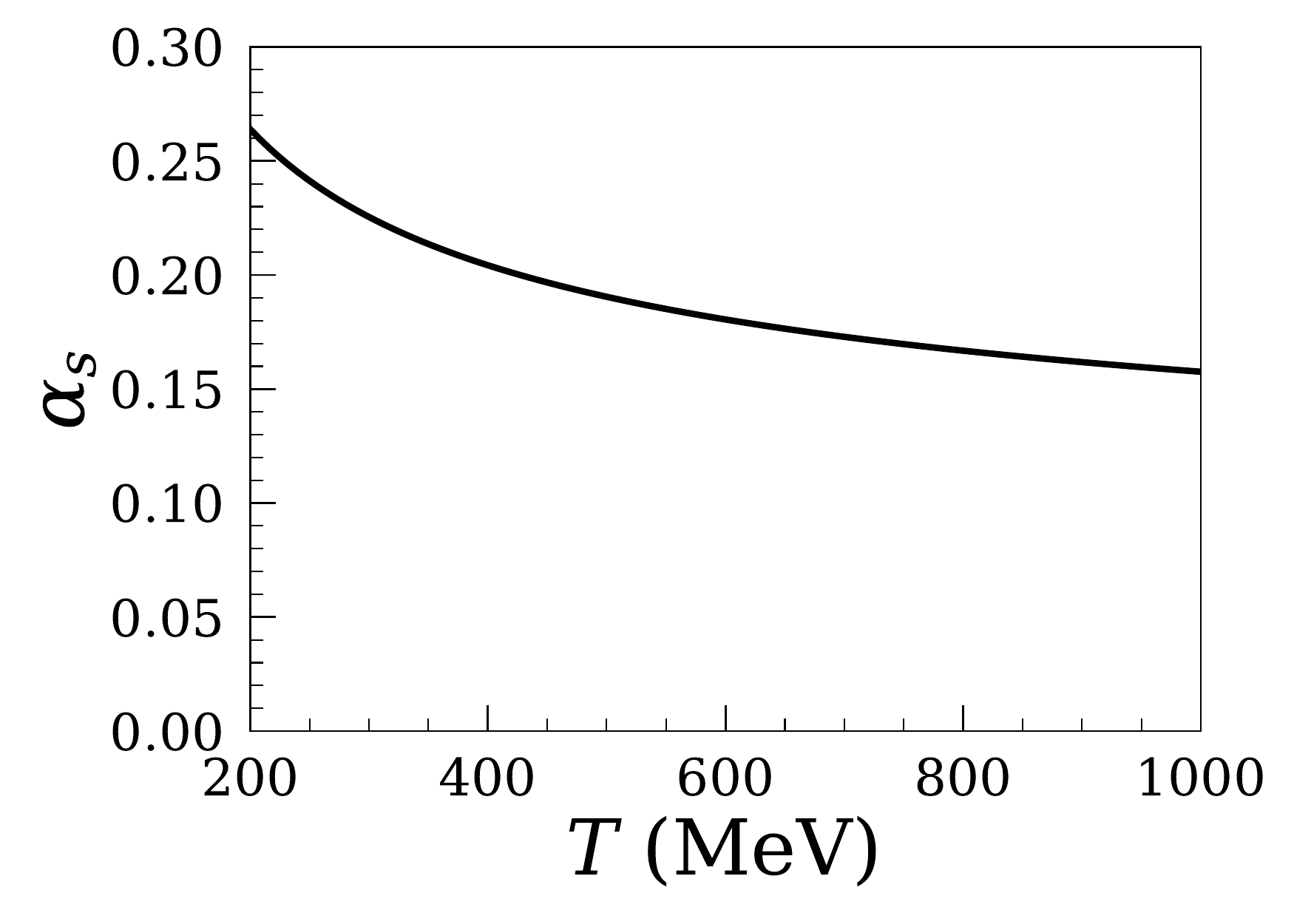}}
\caption{The strong coupling $\alpha_s$ inferred from a comparison of perturbative QCD to lattice results from \cite{Albright1}.}
\label{alphas}
\end{figure}

Figure 2 shows the results from real time lattice Hamiltonian simulations \cite{Moore1}.  The error bars represent statistical uncertainties only.  The systematic uncertainties are unknown but expected to be much larger.  Results for $\alpha_s$ greater than 0.093 could not be extracted from those simulations.  The weak coupling formula from Eq. (\ref{weak}) is shown for comparison.  It represents the numerical results very well for the smallest value of $\alpha_s$, but increasingly deviates with stronger coupling.  The result from AdS/CFT Eq. (\ref{AdSCFT}) is also shown.

\begin{figure}[H]
\center{\includegraphics[width=290pt]{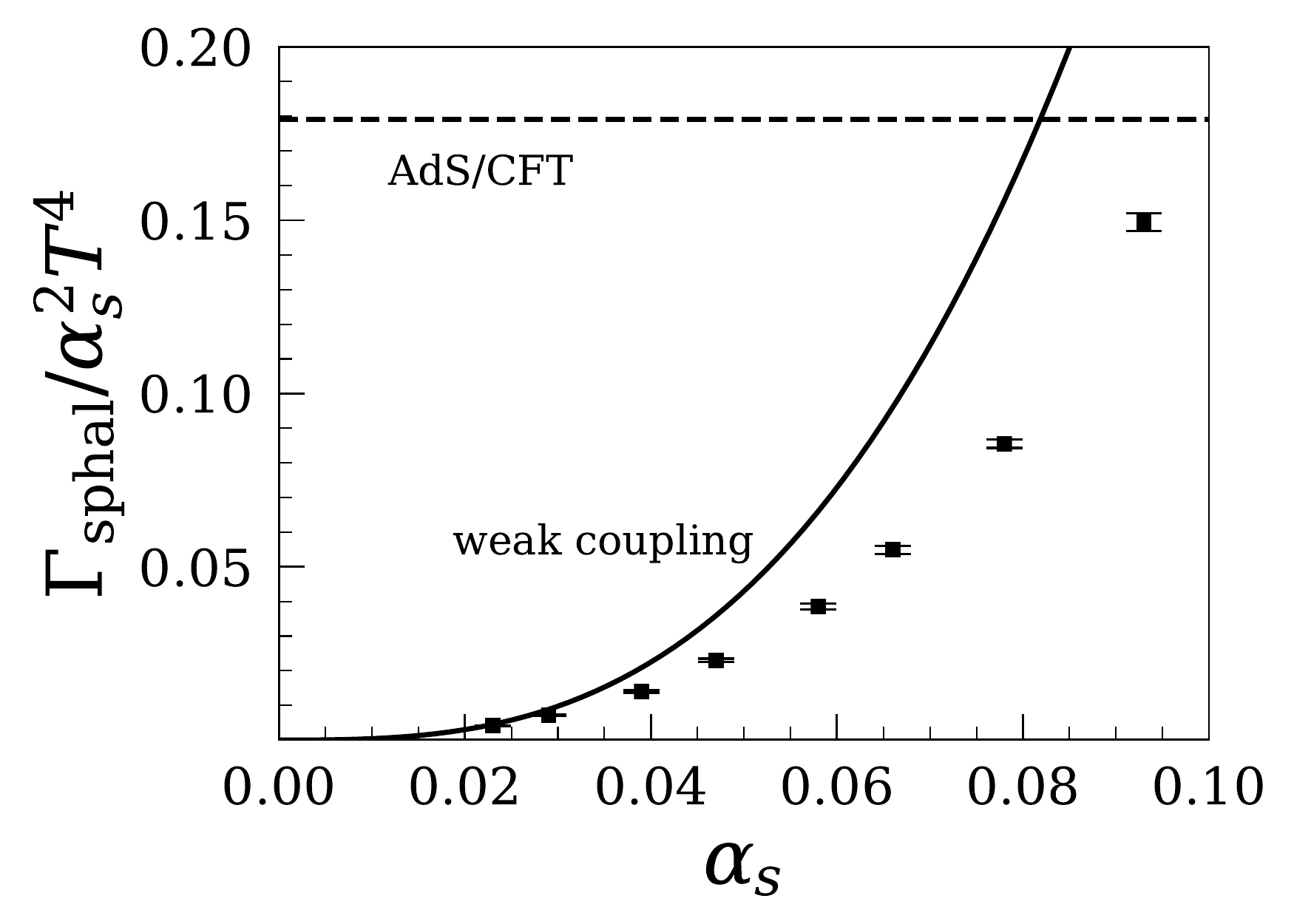}}
\caption{Comparison of sphaleron rates from Eqs. (\ref{weak}) and (\ref{AdSCFT}) and from real time lattice Hamiltonian simulations (points) from \cite{Moore1}.  Error bars represent statistical uncertainties only.}
\label{comparerates}
\end{figure}

Equation (\ref{weak}) necessarily requires a choice of normalization scale.  It was chosen according to the principle of the ``fastest apparent convergence".  This resulted in the number 3.041 appearing in that formula, which can be combined with the logarithm for that choice of scale.  We changed the aforementioned number to 1.26 to better represent the numerical results.  It simply represents a different choice of the normalization scale.  In principle physical results should be independent of the choice, but to any finite order in perturbation theory it does make a quantitative difference.  The change is shown in Figure 3.  Although it does not represent the numerical results quite as well for the smallest value of $\alpha_s$, it does a much better job of representing them over the full range.  Therefore we shall use it to extrapolate the numerical results to the larger values of $\alpha_s$ relevant for heavy ion collisions.

\begin{figure}[H]
\center{\includegraphics[width=290pt]{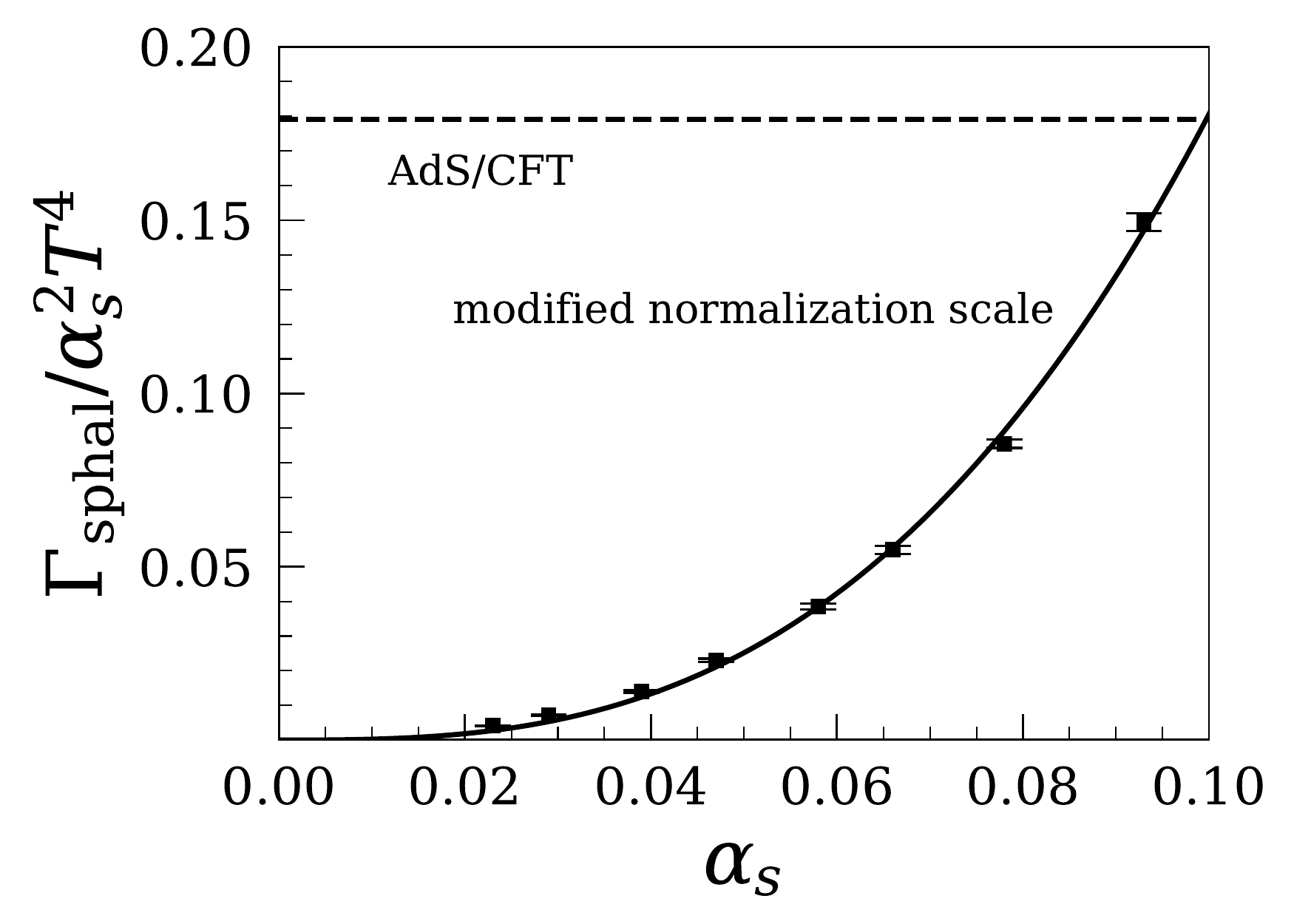}}
\caption{Comparison of sphaleron rates from Eq. (\ref{weak}) using a modified normalization scale and from real time lattice Hamiltonian simulations (points) from \cite{Moore1}.  Error bars represent statistical uncertainties only.}
\label{comparemodrates}
\end{figure}

Figure 4 shows the microscopic equilibration time from Eq. (\ref{micro}) divided by the two estimates for the axial charge equilibration time. As discussed above, this ratio should be much less than 1 for the sphaleron transition rate calculations to be reliable. Given the systematic uncertainties in all quantities presented the results from the weak coupling calculation, with the modified normalization scale as adjusted to match the semi-classical simulations, suggests that those results are on the limit of applicability.  On the other hand, the ratio test is quite well satisfied when using the sphaleron rates from AdS/CFT.
 
\begin{figure}[H]
\center{\includegraphics[width=290pt]{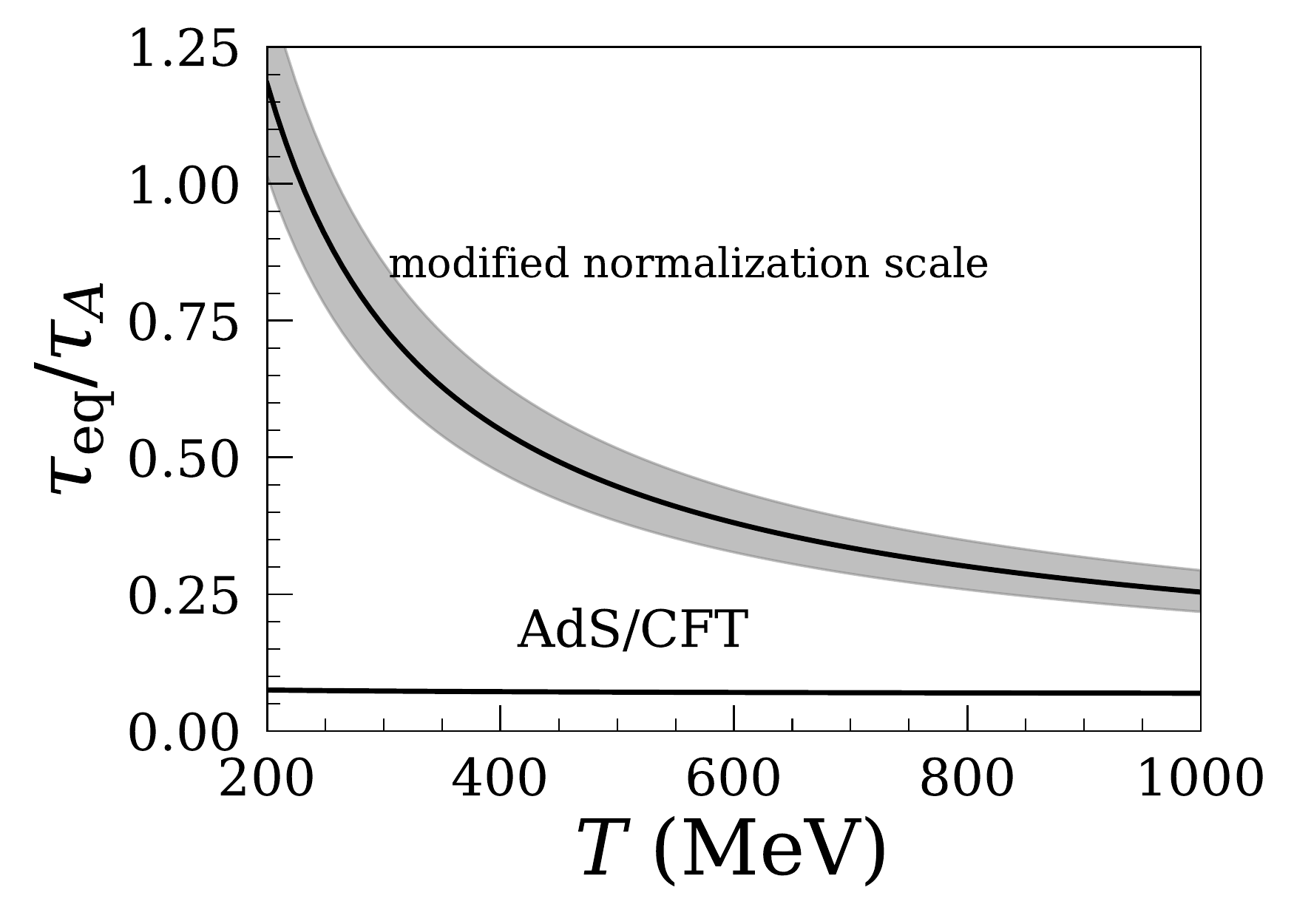}}
\caption{Ratio of microscopic Eq. (\ref{micro}) to axial charge relaxation times.  For the classical calculation of the sphaleron rate to be valid requires this ratio to be much less than one.  The gray bands represent a $\pm$5\% variation in $\alpha_s$.}
\label{tau_ratio}
\end{figure}

Figure 5 shows the axial charge equilibration time in units of fm/c for the two choices of sphaleron rate.  Given that temperatures in the QCD phase in heavy ion collisions at RHIC and LHC range between 200 and 600 MeV, and expansion time scales range between 1 and 10 fm/c, the likelihood for axial charge to be in equilibrium based on the AdS/CFT rate is marginal at best.  The rate extrapolated from weak coupling with the modified normalization scale would appear more likely to keep axial charge in equilibrium.  However, based on Figure 4 the reliability of the calculation of that rate is questionable.
 
\begin{figure}[H]
\center{\includegraphics[width=290pt]{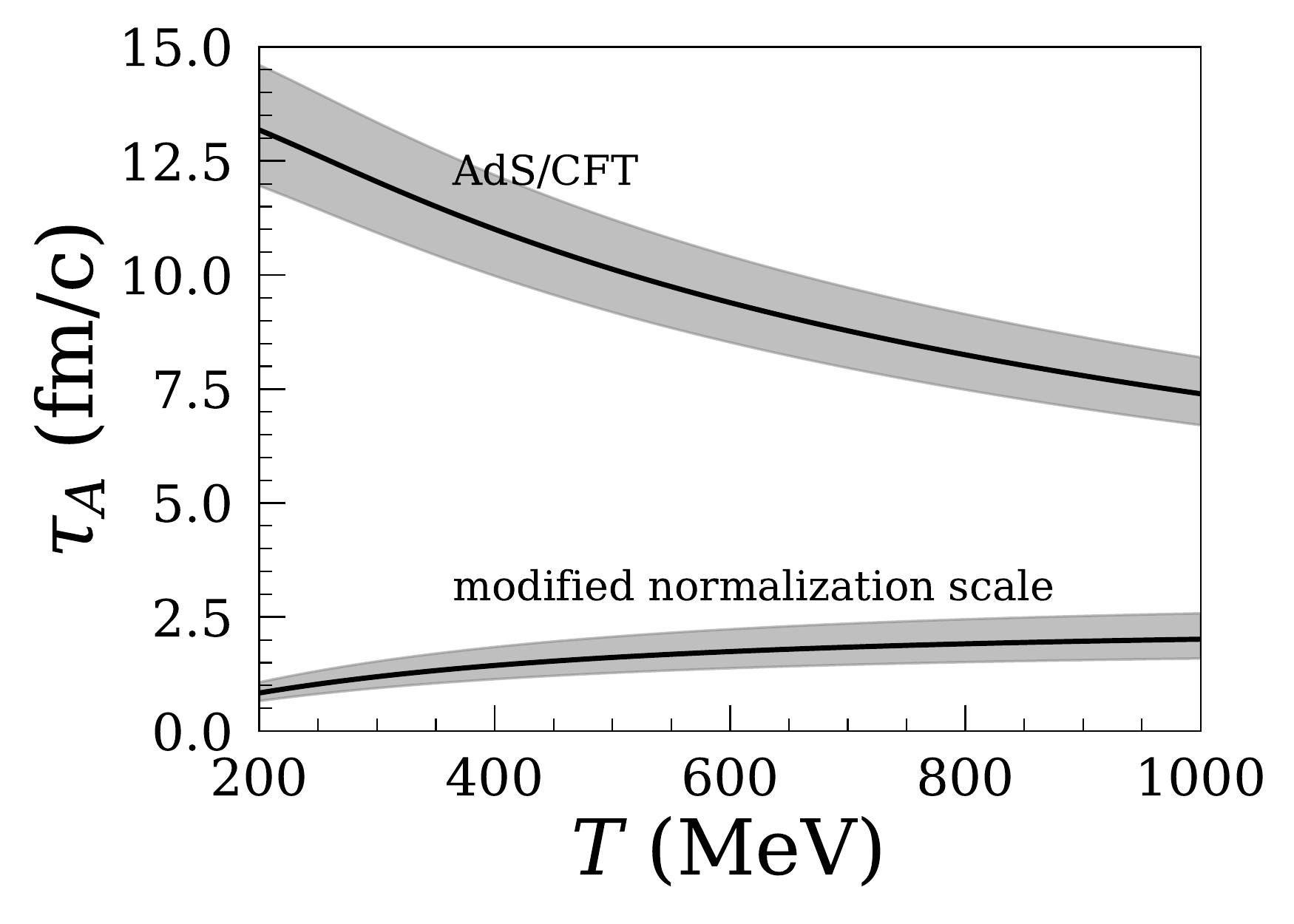}}
\caption{The axial charge relaxation time $\tau_A$ using the modified normalization scale mentioned in the text and from AdS/CFT Eq. (\ref{AdSCFT}).  The gray bands represent a $\pm$5\% variation in $\alpha_s$.}
\label{tau5}
\end{figure}

\section{Nonzero Quark Masses}

Very little is known about the effect of quark masses on the sphaleron transition rate.  It is a difficult problem, and here we briefly comment on the calculations reported in Ref. \cite{Mace1}.  A gauge field configuration was chosen to mimic a sphaleron transition between topologically distinct classical vacua.  Then, the Dirac equation was solved using Hamiltonian dynamics in real time on a spatial lattice.  Backreaction of the quarks on the gluon field was not included.  Rather, the study was focused on the charge separation effect (CSE).  Spatial and temporal coordinates were normalized in units referred to as $r_{\rm sphal}$ and $t_{\rm sphal}$, respectively, with the fixed ratio $r_{\rm sphal}/t_{\rm sphal} = 2/3$.  The CSE was maximal for (nearly) massless quarks and decreased with increasing mass.  When 
$m t_{\rm sphal}$ became of order 1 the effect essentially disappeared.

The current quark masses are $m_u \approx 4$ MeV, $m_d \approx 7$ MeV, and $m_s \approx 110$ MeV.  Even the strange quark mass is small enough compared to the temperature to be considered essentially massless with respect to the equation of state and the Debye screening mass.  However, in the present context the relevant scale is $t_{\rm sphal}$, not $T$.  ${\it If}$ we can identify $t_{\rm sphal}$ with $\tau_A$ {\it then} Fig. 5 would suggest that the strange quark would nearly or entirely wipe out the CSE.  But could the CSE be saved by the very small up and down quark masses?  Not enough is known about the effect of quark masses on the sphaleron rate to answer this question.

What is the relationship between chirality flip and helicity flip?  For massless quarks chirality and helicity are the same while they are opposite for anti-quarks.  For massive quarks the relationship is very subtle and dependent on the physical observable.  With the notation $R$ and $L$ for right and left chirality, $+$ and $-$ for positive and negative helicity, and a bar denoting antiparticles, the relations for massless quarks are $N_R = N_+$, $\bar{N}_R = \bar{N}_-$, $N_L = N_-$, and $\bar{N}_L = \bar{N}_+$.  Therefore
\be
Q_A = (N_R - \bar{N}_R) - (N_L - \bar{N}_L) = (N_+ + \bar{N}_+) - (N_- + \bar{N}_-)
\ee
This shows that the helicity and chirality equilibration times are equal for massless quarks.  On the other hand, it says nothing about the momentum dependence of the helicity flip rate which is important for heavy ion collisions.

\section{Conclusion}

The CME is an interesting phenomenon currently being investigated in
high-energy heavy ion collisions. However, unlike spin polarization,
there is currently no experimental evidence found for it. In this work
we focused our attention on the axial charge relaxation time.  We
began by rederiving the axial charge equilibration time as a function
of the sphaleron transition rate at finite temperature for massless
quarks. This rate had been computed with different approaches
depending on the value of the strong coupling. In the weak coupling
limit we employed a previously derived semi-analytical expression fit
to real time lattice Hamiltonian calculations. In the opposite limit,
through the AdS/CFT correspondence, we used the rate as calculated in
the ${\cal N} = 4$ Super Yang-Mills theory. To assess the applicability of
these sphaleron rate calculations we compared the respective
equilibration times to the microscopic thermalization time. We found
that in the weak coupling limit the result is marginally valid, while
in the strong coupling AdS/CFT limit the validity appears more
certain.  On the other hand, the strong coupling AdS/CFT associated
relaxation time seems to be too long compared to the expansion time to allow for the CSE, while the
weak coupling associated relaxation time seems to allow for the CSE up
to the time of hadronization.  The role played by the quark masses is
uncertain and remains to be accurately determined.  At this time we
can reach no firm conclusions.

Note Added: Recently the STAR Collaboration at RHIC carried out a blind analysis of isobar collisions at $\sqrt{s_{NN}} = 200$ GeV in a search for the CME \cite{ThirdSTAR}.  They found that the CME background is different between the two isobar species but no CME signature that satisfies their predefined criteria was observed.

\section*{Acknowledgement}
We thank L. D. McLerran for suggesting this investigation during a presentation by J. I. K. hosted by the Institute for Nuclear Theory Program INT-20-1c on Criticality and Chirality in May 2020 and for comments on the manuscript.  The work of J. I. K. was supported by the U.S. Department of Energy Grant DE-FG02-87ER40328.  The work of E. R. was supported by the U.S. National Science Foundation Grant PHY-1630782 and by the Heising-Simons Foundation Grant 2017-228.

\end{document}